\documentclass[twocolumn,pre,showpacs,amsmath,amssymb]{revtex4-1}
\usepackage{geometry}                
\usepackage{graphicx}
\usepackage{amssymb}
\usepackage{epstopdf}
\usepackage{shadow}
\newcommand{\be}{\begin{eqnarray}}
\newcommand{\ee}{\end{eqnarray}}
\newcommand{\la}{\langle}
\newcommand{\ra}{\rangle}
\newcommand{\tr}{{\rm Tr\,}}

\begin{document}
\title{Thermodynamics of Evolutionary Games}
\author{Christoph Adami}
\email{adami@msu.edu}
\affiliation{Department of Microbiology \& Molecular Genetics}
\affiliation{Department of Physics \& Astronomy}
\affiliation{BEACON Centre for the Study of Evolution in Action\\
Michigan State University, East Lansing, MI 48824}
\author{Arend Hintze}
\affiliation{Department of Computer Science \& Engineering}
\affiliation{Department of Integrative Biology}
\affiliation{BEACON Centre for the Study of Evolution in Action\\
Michigan State University, East Lansing, MI 48824}
\begin{abstract}
How cooperation can evolve between players is an unsolved problem of biology. Here
we use Hamiltonian dynamics of models of the Ising type to describe populations of cooperating and defecting players to show that the equilibrium fraction of cooperators is given by the expectation value of a thermal observable akin to a magnetization. We apply the formalism to the Public Goods game with three players, and show that a phase transition between cooperation and defection occurs that is equivalent to a transition in one-dimensional Ising crystals with long-range interactions. We then investigate the effect of punishment on cooperation and find that punishment plays the role of a magnetic field that leads to an ``alignment'' between players, thus encouraging cooperation. We suggest that a thermal Hamiltonian picture of the evolution of cooperation can generate other insights about the dynamics of evolving groups by mining the rich literature of critical dynamics in low-dimensional spin systems.
\end{abstract}

\maketitle
\mbox{}

Cooperation is a particularly interesting phenomenon in the context of evolution. Evolution acts on short-term benefits, which makes cooperators vulnerable to exploitation in the form of cheating or ``defection'' even if cooperation is a strategy with higher payoffs in the long-term, creating what is known as the ``dilemma of cooperation''.  It is often stated that because of the dilemma, the expected outcome of evolution should be defection, rendering the plethora of examples for cooperators in nature mysterious. However, there are number of different mechanisms that nevertheless enable cooperation
~\cite{MaynardSmith1982,Axelrod1984,HofbauerSigmund1998,Nowak2006} suggesting that, contrary to the naive expectation, cooperation is after all the natural outcome of evolution when mechanisms enabling assortment (such as discrimination via communication) are available~\cite{Adamietal2016a}. These results have been obtained 
using mathematics as well as computational-simulation tools. The mathematical results in particular provide insight into the evolutionary dynamics giving rise to cooperation from inspecting closed-form solutions, but such solutions are hard to come by when populations are finite, are not well-mixed, or are subject to significant mutation~\cite{Adamietal2016a}. Recently, progress was made in understanding the evolutionary dynamics on games played on arbitrary grids~\cite{Allenetal2017}, but closed-form solutions predicting the ``critical point'' for the transition between cooperation and defection still do not exist. Here, we use methods borrowed from statistical physics that show the path to such general formul\ae. 

Prior investigations of several standard evolutionary games ~\cite{SzaboFath2007,Szolnokietal2009,Iliopoulosetal2010,AdamiSchossauHintze2012,HintzeAdami2015,SzaboBorsos2016} 
revealed that the evolutionary process often critically depends on a single parameter that causes an abrupt change in winning strategy. In some cases it is possible to move the parameter beyond the critical point without triggering the transition---the hallmark of hysteresis~\cite{HintzeAdami2015}. These results suggest that there is an underlying analogy between evolutionary game dynamics and the statistical description of phase transitions. Indeed, Szab\'o and Hauert~\cite{SzaboHauert2002,HauertSzabo2005} applied mathematical methods that are used to describe critical phase transitions like the ones found in the celebrated Ising model~\cite{Ising1925} to evolutionary games on a lattice, and showed (via numerical simulation, as well as the pair-approximation on square lattices) 
that the Prisoner's Dilemma (PD) game dynamics on random regular lattices fall into the directed percolation class of phase transitions. 

Here we take a different approach, by explicitly constructing Hamiltonians for game dynamics inspired by Ising-type models, and studying games on finite regular lattices analytically (albeit only in one dimension).  It might at first appear odd to consider thermal game theory, as temperature plays no role in evolutionary dynamics. In physics, thermal effects are due to fluctuations in energy, but payoffs in evolutionary games can fluctuate as well, for a number of different reasons. For example, a finite evolving population is subject to drift and thus to a random element in the payoffs. Mutations that change strategies can play a similar role. In evolutionary games, we can summarize the effect of fluctuations by introducing a parameter that controls the {\em strength of selection} in the game, using the ``strategy adoption'' mode of selection (see~\cite{HauertSzabo2005} and below). While the dynamics under this rule is not precisely the same as the ``strategy inheritance'' mode of Darwinian selection, the differences (also discussed in~\cite{HauertSzabo2005}) are irrelevant for our purposes. The relationship between game dynamics and Ising-type models has been reviewed recently~\cite{SzaboBorsos2016}

To introduce our method and notation, we first study the Prisoner's  Dilemma Hamiltonian at finite temperature and recover well-known results. We then apply the method to the Public Goods game without punishment, which turns out to be equivalent to an Ising model with long-range interactions, but without a magnetic field. We then add punishment to the Public Goods game, leading to an Ising model with magnetic field (and corresponding hysteresis effects) that we solve exactly. 

\subsection{Prisoner's Dilemma} 
The Prisoner's Dilemma is a game played between two individuals, in which both players have to make a decision about whether to cooperate or to defect. After both players have made their choice--to cooperate (C) or to defect (D)--their actions are revealed and players receive a payoff according to a payoff matrix (note that the values in the matrix correspond to the payoff given to the ``row" player)
\be
E=\bordermatrix{\mbox{} & C & D \cr
C & R & S \cr
D & T & P }\;.
\label{RSTPtable}
\ee
The payoffs in that matrix define the type of game to be played. To obtain a Prisoner's Dilemma, we must have~\cite{Axelrod1984} $T>R>P>S$. If the game is played repeatedly it becomes the {\em iterated} Prisoner's Dilemma (IPD), a variant not considered here. Evolutionary game theory focuses on determining what strategies are evolutionarily stable in a population of strategies. In the simplest case, competition is between two unconditional deterministic strategies: one that always cooperates and one that always defects. A population starts out as a mix of both strategies, and players interact with a defined number of neighbors. Each player's performance is evaluated by accumulating all payoffs received in that round. To model evolution, randomly-picked players (called focal players) can now either maintain their strategy or adopt the strategy of a competitor. Over time this process will lead to the spread of successful strategies and thus to evolution. This process of probabilistic strategy adoption is similar to the dynamics of strongly interacting spins described by Glauber~\cite{Glauber1963}. In such a model of ferromagnetism, adjacent particles interact so that their spins will predominantly align (a spin adopting the state of its neighbor), giving rise to an overall magnetization that depends on the temperature of the system. In the following, we explore this analogy more deeply.

We first derive the thermodynamics of the Prisoner's Dilemma with a payoff matrix where we set the reward $R=b-c$ (the benefit of cooperation minus the cost), while the temptation payoff $T=b$ (obtaining the benefit without bearing the cost). At the same time, the so-called ``sucker-payoff'' $S=-c$ due to paying the cost without any benefit, while $P=0$ is the ``punishment'' for both players mistrusting each other. In all of the following, we assume $c\geq0$ as well as $b-c\geq0$, so that the net benefit $r=b-c\geq0$, ensuring that a dilemma exists. Indeed, even though the benefit outweighs the cost ($r>0$), the Nash equilibrium and evolutionarily stable strategy is known to be defection, not cooperation. The payoff matrix in terms of these values then becomes
$
E=\left(
\begin{array}{cc}b-c & - c   \\
 b & 0
\end{array} \right)
$.     

To define a Hamiltonian (an operator that describes the total energy for this system) we can transform the payoffs into an energy by subtracting the payoff from its largest possible value. However, as this only adds a global constant it will cancel in observables, so to understand the population dynamics in terms of thermodynamics we can keep the payoff as is. A Hamiltonian is an operator that acts on a vector space (Hilbert space).  A basis for the Hilbert space is spanned by the cooperative strategy C and the defecting strategy D by the vectors
$C=|0\ra =\left(\begin{array}{c} 1  \cr 0 \end{array}\right)$ and 
$ D=|1\ra =\left(\begin{array}{c} 0  \cr 1 \end{array}\right)$.

In analogy to Ising spin systems, the Hamiltonian for the PD game can then be written in terms of the energy matrix $E$ and the projectors
$
P_0=|0\ra\la 0|=\left(
\begin{array}{cc}1 & 0  \\
0 & 0 
\end{array} \right)$ and $ P_1=|1\ra\la 1|=\left(
\begin{array}{cc}0 & 0  \\
0 & 1
\end{array} \right)
$
as
\be
H=\sum_{i=1}^N \sum_{m,n=0}^1 E_{mn}P_m^{(i)}\otimes P_n^{(i+1)}\;,
\ee
where the sum over $i$ goes over all the sites in this one-dimensional ``spin chain".

We proceed by calculating the thermal partition function of the system by writing ($\beta=1/T$ is the inverse of the temperature, which the reader will not confuse with the temptation payoff)
\be
Z=\tr  e^{-\beta H}&=&\sum_{x}\la x|e^{-\beta H}|x\ra \;,
\ee
where $|x\ra=|m_1m_2\cdots m_N\ra$ is a circular chain so that the $N$th site is adjacent to the first site. It is then easy to see that
\be
Z&=&\sum_{m_1\cdots m_N} e^{-\beta(E_{m_1m_2} +E_{m_2m_3}+\cdots+E_{m_Nm_1})}\nonumber\\
&=& \sum_{m_1\cdots m_N} U_{m_1m_2}U_{m_2m_3}\cdots U_{m_Nm_1}\nonumber \\
&=&\tr U^N\;,
\ee
where $U_{ij}=e^{-\beta E_{ij}}$.

\begin{figure}[htbp] 
   \centering
   \includegraphics[width=3in]{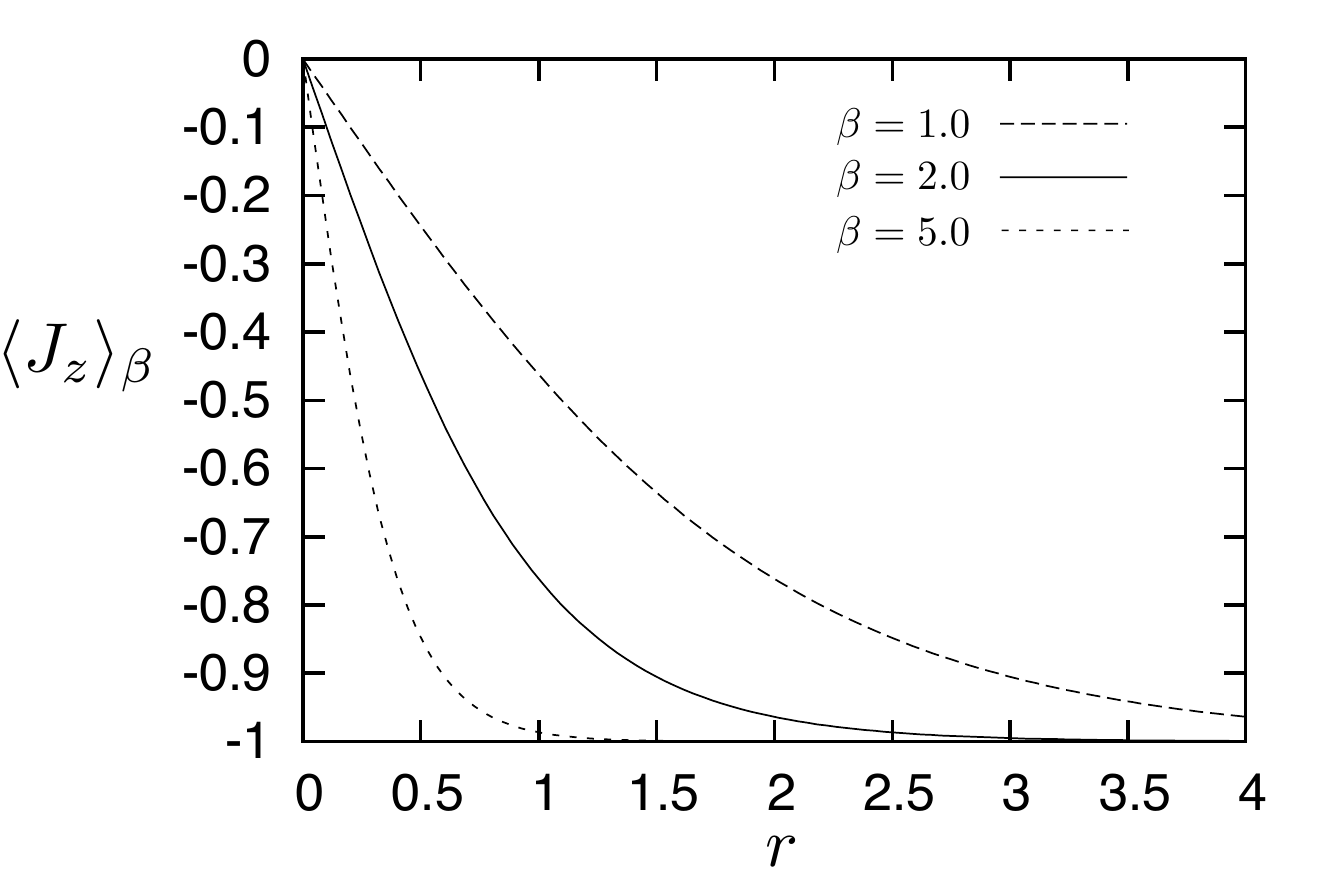} 
   \caption{Order parameter $\la J_z\ra_\beta$ as a function of the net reward $r=b-c$, for three different temperatures. As opposed to the game in two dimensions~\cite{HauertSzabo2005}, the phase transition occurs at $r=0$.}
   \label{Ising-PD}
\end{figure}

To determine the equilibrium population composition, we define an {\em order parameter} given by the fraction of cooperators minus the fraction of defectors. For spin chains this is equal to the {\em magnetization} of the chain, defined using a spin operator $J_z$ for which 
$
\la 0|J_z|0\ra = 1$ and 
$\la 1|J_z|1\ra = -1$.
This can be achieved, e.g., with  ($\sigma_z$ is a Pauli matrix)
\be J_z= \sigma_z=P_0-P_1\;.  \label{spin}
\ee
We will understand this operator to act on the ``row'' player (that is, the first spin of the pair). For a chain of length $N$, 
\be
J_z=\sum_{i}^N(P_0^{(i)}-P_1^{(i)})\;, \label{spin-sum}
\ee
so that 
\be
&\sum_x&\la x| \hat J_ze^{-\beta H}|x\ra=N\tr (U'U^{N-1})
\ee
due to the cyclic property of the trace. Here we introduced the matrix $U'_{ij}=(-1)^iU_{ij}$.
An explicit calculation shows that (recall that $r=b-c$) 
\be
Z=\tr U^N= (1+e^{-\beta r})^N\;,
\ee
while since $U'U=(1+e^{-\beta r})U'$ and $\tr U'=-1+e^{-\beta r}$
\be
\tr (U'U^{N-1})=(1+e^{-\beta r})^{N-1}(-1+e^{-\beta r})\;,\ \ \ \ \ \ \ 
\ee
so that finally the thermal expectation value of the magnetization is
\be
\la J_Z\ra_\beta &=&\frac1 Z\sum_x\la x| \hat J_ze^{-\beta H}|x\ra\nonumber \\
&=&-N\tanh(\beta r/2)\;.
\label{orderpd}
\ee
We show the magnetization {\em per player} [Eq.~(\ref{orderpd}) divided by $N$]  as a function of the critical parameter $r$ in Fig.~\ref{Ising-PD}, and see that at low temperatures (high $\beta$) the population will consist mostly of defectors (negative magnetization) as this is the Nash equilibrium. We note that the parameter $r$ plays the same role as the interaction strength $J$ in the standard Ising model.
The phase transition (vanishing magnetization) occurs at $r=0$ (the ``boundary'' of the parameter values), which is expected from the general arguments of van Hove~\cite{vanHove1950} and of Landau~\cite{LandauLifschitz1987} that forbid phase transitions in one-dimensional systems. Thus, we do not observe cooperation in the one-dimensional Prisoner's dilemma, as is of course well-known.

\subsection{Public Goods game in one dimension}
The PD game we just described turns out to be the two-player version of the more general Public Goods (PG) game. 
The PG game is a staple of evolutionary game theory as well as experimental economics~\cite{Olson1971,DavisHolt1993,Ledyard1995}, and has been used to understand the {\em Tragedy of the Commons}~\cite{Hardin1968}, a social dilemma that can lead to the overuse of public resources (for example, overfishing) because of selfish behavior.  In the PG game, payoffs are defined for cooperators and defectors via 
\be
\Pi_C&=&\frac r{(k+1)}(N_C+1)-1 \label{pg1}\\
\Pi_D&=&\frac{rN_C}{(k+1)} \label{pg2}
\ee
where $\Pi_C$ is the payoff for a cooperator ($\Pi_D$ for a defector). $N_C$ is the number of cooperators in the neighborhood (not counting the focal player, so it is the number of cooperators in the player's periphery), and $r$ is the reward multiplier (synergy factor). These are the rules for a game with $k+1$ players in a group. In the following, we will treat the game in one dimension (so $k=2$). 

The rules (\ref{pg1}-\ref{pg2}) imply a payoff matrix
\be
\Pi_{\rm C}=\bordermatrix{\mbox{} & {\rm C} &{\rm D}\cr
{\rm C} & r-1  & \frac23r-1\cr 
{\rm D} &  \frac23r-1 & \frac13r-1}
\ee
for cooperators, where the matrix elements indicate the states of spins in the periphery of the focal player. For example,  $r-1$ is the payoff for a cooperator surrounded by two cooperators. The payoff matrix for defectors is simply
$
\Pi_{\rm D}=\Pi_{\rm C}-(\frac13 r-1)\;.
$.

\begin{figure}[htbp] 
   \centering
   \includegraphics[width=2.5in]{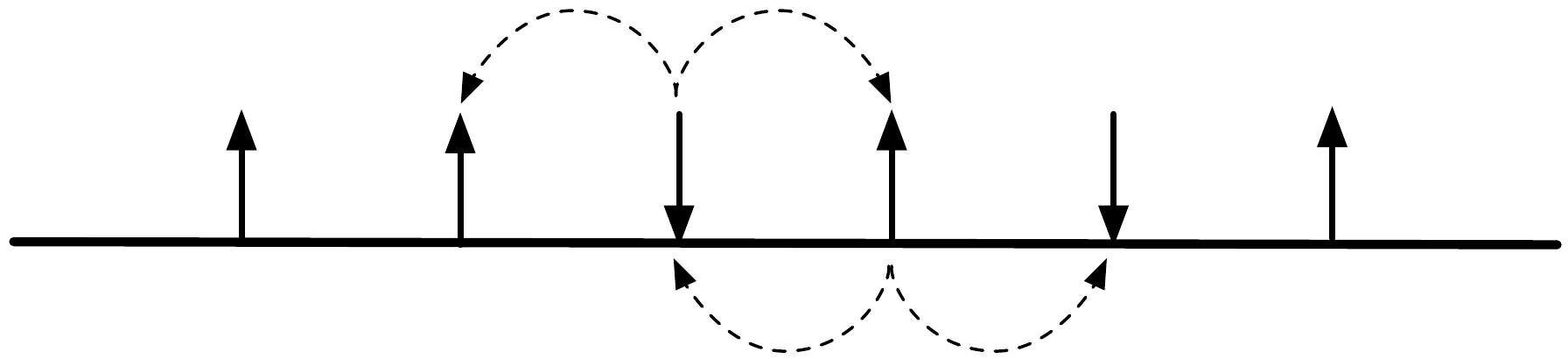} 
   \caption{One-dimensional string of players in a Public Goods game that interact with two nearest-neighbors. Because players interact with more than one nearest-neighbor, effectively next-to-nearest neighbors interact.}
   \label{fig:PG-string}
\end{figure}

We now construct a Hamiltonian to solve this evolutionary model exactly in two cases: one where the dynamics maximize the mean payoff of the population, and one in which the payoff of an individual is maximized. Naturally, we expect a correspondence with the evolutionary scenario only in the latter case. In this one-dimensional game, the population is arranged linearly so that each player forms a group with its left and right neighbor $(k=2)$, see Fig.~\ref{fig:PG-string}. 

As mentioned earlier, we can create matrices for energies that should be minimized (rather than payoffs that need to be maximized)
by subtracting the payoffs from the maximal payoff (here, $r-1$), leading to a ground state that has zero energy. Strictly speaking, the Hamiltonian for this system should be written as an interaction of three spins, but we will often write it in terms of a two-spin interaction matrix {\em conditional} on the state of the focal spin. For example, we can write 
\be
E^{(C)}=\left(\begin{array}{cc}
                       0  & \frac13r \\
                          \frac13r &  \frac23r \end{array}\right),
                           \; E^{(D)}=\frac13r-1+E^{(C)}.
                           \;\;\;\;\; \;\label{pay3}
\ee
We write a Hamiltonian for cooperators using these energies and the projectors previously defined 
\be
H_C^{(i)}=\sum_{i=0}^N\sum_{m,n=0}^1 E^{(C)}_{mn}P_m^{i-1}\otimes P_n^{i+1}\;, \label{hamc}
\ee
and similarly for $H_D^{(i)}$. 
The total Hamiltonian is (recall that $P_0$ projects onto a cooperator, so that $P_0|0\ra=| 0\ra$ while $P_0|1\ra=0$)
\be
H= \sum_{i=1}^N H_C^{(i)}P^{(i)}_0+H_D^{(i)}P^{(i)}_1\;.
\ee
Using the spin operator (\ref{spin-sum}) and the methods outlined earlier, we obtain after a somewhat tedious calculation
\be
\la J_z\ra_\beta= \frac1Z \tr (J_z e^{-\beta H})=N \tanh \frac\beta2(r-1)\;,\;\;\; \label{wrong-t}\ \ \ 
\ee
suggesting a phase transition at $r=1$, in contradiction with the standard expectation~\cite{HintzeAdami2015} that suggests a transition at $r=3$ (see below). The reason for this discrepancy is not difficult to find: Hamiltonian dynamics minimize the energy of the entire spin chain, which is equivalent to maximizing population fitness as a whole. Darwinian evolution, however, does not optimize population fitness, but rather maximizes the fitness of a single individual {\em within} a population.

\begin{figure}[htbp] 
   \centering
   \includegraphics[width=3in]{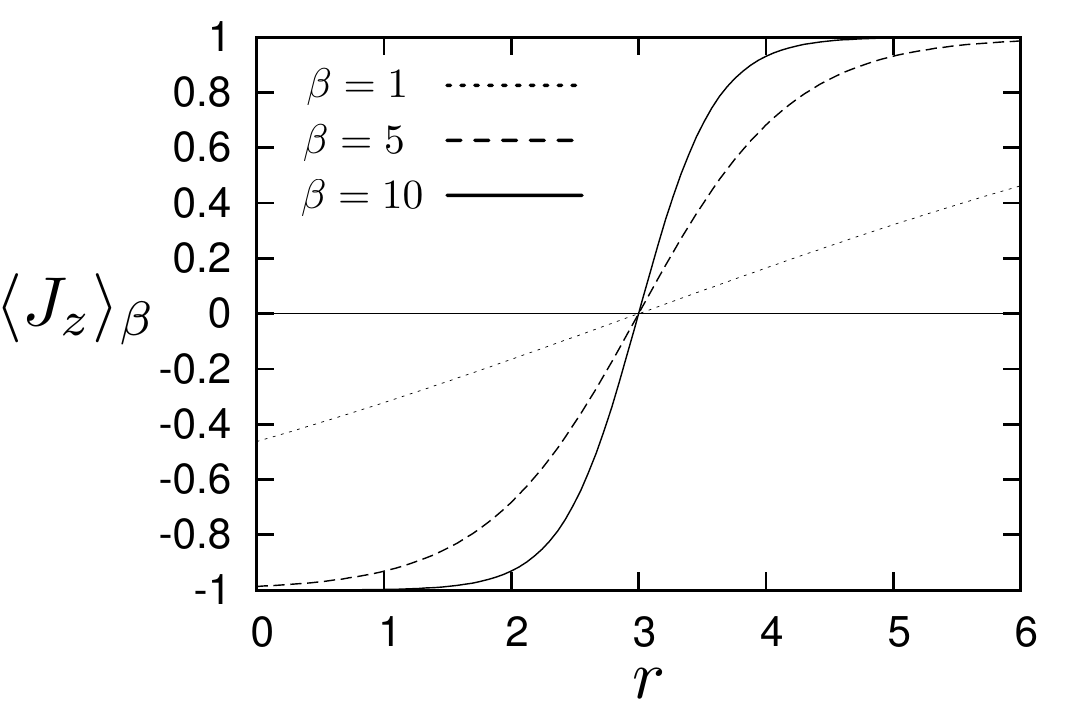} 
   \caption{Exact solution for the order parameter $\la J_z\ra_\beta$ as a function of synergy parameter $r$ for three different temperatures, from Hamiltonian dynamics.}
   \label{ising}
\end{figure}

We can implement the latter dynamic by dropping the sum over sites in Eq.~(\ref{hamc}), and consider only the contribution to the energy from a single spin with its two neighbors. In that case (we take the middle site to be the focal site whose energy is minimized) 
\be
Z&=&\sum_{m_1m_2m_3}\la m_1m_2m_3| e^{-\beta H_{m_2}}|m_1m_2m_3\ra\nonumber \\
&=& \sum_{m_1m_3} (U_{m_1m_3}+V_{m_1m3})
\ee
where $U$ is the ``cooperative" matrix $U=e^{-\beta H_0}$                
while the defector matrix $V=e^{-\beta H_1}=e^{\beta(r/3-1)}U$ because defector energies differ by $r/3-1$ from cooperator energies, see Eq.~(\ref{pay3}).  
Then,
\be
Z&=&\sum_{m_1m_3}U_{m_1m_3}(1+e^{-\beta({r/3-1})})\nonumber \\
&=&(1+e^{-\beta\frac r3})^2(1+e^{-\beta(\frac r3-1)})\;.
\ee
Using the spin operator defined in Eq.~(\ref{spin}) 
we obtain (again for a single focal player in the middle position)
\be
&&\sum_{m_1m_2m_3}\la m_1m_2m_3|J_ze^{-\beta H_{m_2}}|m_1m_2m_3\ra\nonumber\\
&=& (P_0-P_1)\sum_{m_1m_3}(P_0U_{m_1m_3}+P_1V_{m_1m_3})\nonumber\\
&=& (1-e^{-\beta(r/3-1)})(1+e^{-\beta\frac r3})^2\;,
\ee
which allows us to calculate the order parameter as
\be
\la J_z\ra_\beta =\frac1Z \tr (e^{-\beta H} J_z)=\tanh \frac\beta2(\frac13r-1)\;. \ \ \ \ \label{order}\ \ \ \ 
\ee
This function is plotted in Figure~\ref{ising}, and suggests that a phase transition with an interior critical point is possible in this game even though the game is one-dimensional, seemingly violating van Hove's theorem~\cite{vanHove1950}. However,  the theorem forbidding internal critical points in one dimension only holds for short-range interactions, while the interaction between three players studied here is not of that kind.

To test the accuracy of our theoretical result, we now simulate the Public Goods game using agent-based methods~\cite{Helbingetal2010,HintzeAdami2015,Adamietal2016a}.
\begin{figure}[htbp] 
   \centering
   \includegraphics[width=3in]{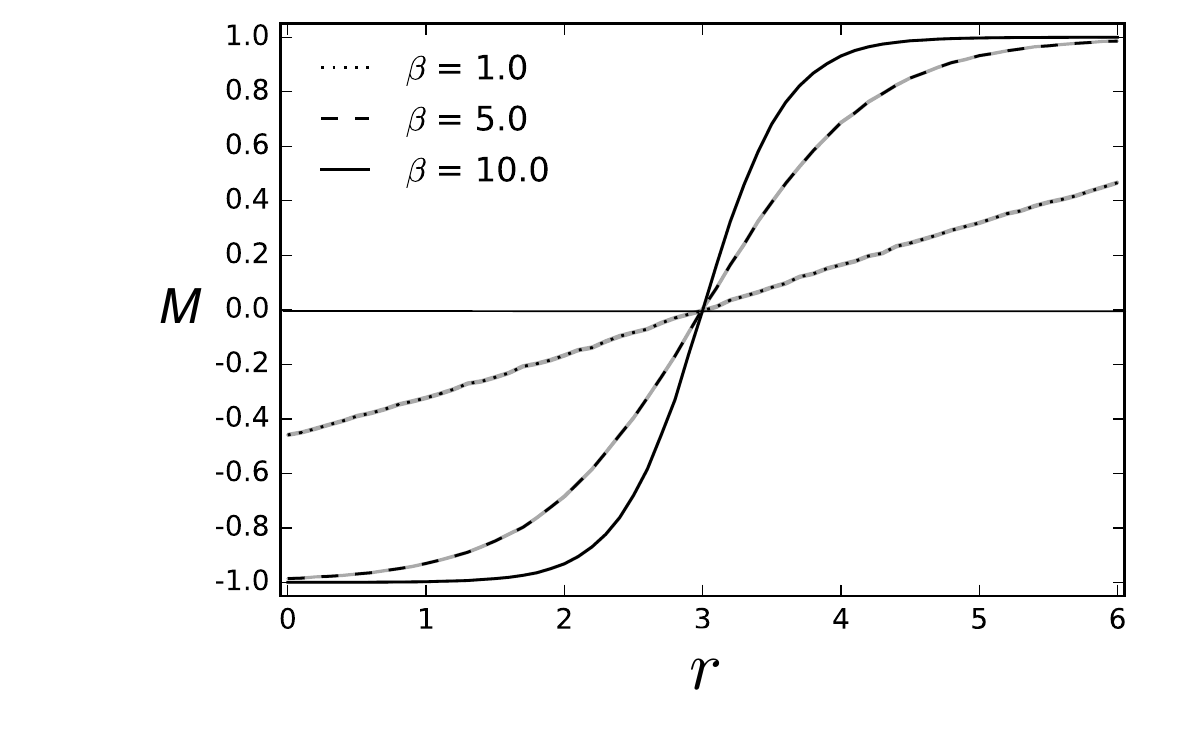} 
   \caption{Fraction of cooperators for a chain of length $2^{10}$, as a function of the synergy parameter $r$ for three different selection strengths defined by $\beta=1/T$. Each data point (each $r$, increments of $\Delta r=0.1$) is the average over 100 replicate agent-based simulations using strategy adoption for $2\times 10^6$ updates. Barely visible grey bands represent standard error.}
   \label{ABS_PGG_noPunish}
\end{figure}
In the agent-based simulations we use a population of 1,024 players that either cooperate or defect, arranged in a one-dimensional chain just as in Fig.~\ref{fig:PG-string}. Which of the two moves an agent chooses is determined by a genome (here a single locus) that evolves. 
At every update, players have a chance to change their strategy by probabilistically adopting the strategy of a competitor (Glauber dynamics, see, e.g.,~\cite{HauertSzabo2005,SzaboFath2007}) using the rule (here $x$ is the focal player while $y$ is an alternative strategy)
\be
p(x\gets y) =\frac{1}{1+e^{-\beta(w_x-w_y)}}\;, \label{strategyAdoptionProbability}
\ee
where $\beta$ is related to the strength of selection and $w$ is the fitness of each player defined by the payoff the player receives. In the case of rejection (i.e., non-adoption) the focal player retains its strategy. 

We define an order-parameter-like function that indicates to what extent the population is in a cooperative or a defective regime. This parameter depends on the fraction of players in the population cooperating ($P_{C}$) and the fraction defecting ($P_{D}$) and is defined as:
\be
M=\frac{P_{C}-P_{D}}{P_{C}+P_{D}} \label{orderParameterPGGsimulationNoPunishment}
\ee
The agent-based simulations confirm that the fate of an evolving population depends critically on the synergy factor $r$ (see Figure~\ref{ABS_PGG_noPunish}), and changes from negative (defection) to positive (cooperation) at $r=3$, in accordance with the critical  $r_c=k+1$ for strategies to evolve cooperative behavior in the Public Goods game~\cite{HintzeAdami2015}. In particular, the simulations confirm the theoretical results with high accuracy.

\subsection{Public Goods game with punishment}
Cooperation evolves in the PG game if the synergy $r$ is at least as large as the group's size $k+1$. However, it is unlikely that in nature cooperation would ever create such a high synergy factor, implying that cooperation cannot evolve in this type of game. It has previously been suggested that punishment is one way to promote cooperation~\cite{FehrGachter2002,FehrFischbacher2003,CamererFehr2006,Sigmundetal2001,Boydetal2003,Brandtetal2003,Helbingetal2010}. By introducing punishment, players can now not only choose between cooperation and defection, but can do this in conjunction with deciding whether or not to punish cheaters. 
This introduces two more strategies: a ``moralist'' M who cooperates and punishes, and an ``immoralist I'' who defects but also punishes~\cite{Helbingetal2010}. 
For every player punished for defecting, each punishing player must pay a cost ($\gamma$), and every player that is punished in such a way suffers a fine ($\epsilon$), thus    
extending the rules (\ref{pg1},\ref{pg2}) to (here, we show the special 1D case $k=2$, for the general case see for example~\cite{HintzeAdami2015})
\be
\Pi_C&=&\frac r3(N_C+N_M+1)-1 \label{pgm1}\\
\Pi_D&=&\frac{r(N_C+N_M)}3 -\epsilon \left(\frac{N_M+N_I}2\right)\label{pgm2}\\
\Pi_M&=&\Pi_C-\gamma \left(\frac{N_D+N_I}2\right) \label{pgm3}\\
\Pi_I&=&\Pi_D-\gamma \left(\frac{N_D+N_I}2\right)\;, \label{pgm4}
\ee
where $N_i$ is the number of players in the immediate neighborhood of the focal player with strategy $i$, $\epsilon$ parameterizes the effect of punishment, while $\gamma$ stands for the cost of punishment (see~\cite{Helbingetal2010,HintzeAdami2015}).

We now study this model thermodynamically, but in order to compare to the evolutionary dynamics we study the regime where the energy of a single site is minimized. To account for the additional strategies (beyond cooperator and defector), we extend the Hilbert space by allowing for a site-dependent magnetization $|i\ra\rightarrow|i\ra|j\ra$,
so that each strategy is defined by a product of spin vectors. If we define punishment as $|0\ra$ and non-punishment as $|1\ra$, we can write the states of the punishing and non-punishing cooperator as 
\be
{\rm M} &=&|0\ra|0\ra=\left(\begin{array}{c} 1  \cr 0 \end{array}\right) \otimes \left(\begin{array}{c} 1  \cr 0 \end{array}\right) = \left(\begin{array}{c} 1  \cr 0 \cr
0 \cr 0 \end{array}\right)\\
{\rm C} &=&|0\ra|1\ra=\left(\begin{array}{c} 1  \cr 0 \end{array}\right) \otimes \left(\begin{array}{c} 0  \cr 1 \end{array}\right) = \left(\begin{array}{c} 0  \cr 1 \cr
0 \cr 0 \end{array}\right)\;.
\ee
The payoffs (\ref{pgm1}-\ref{pgm4}) can be written in terms of a Hamiltonian for each of the four strategies as
\be
H=P_{00} H_M+P_{01} H_C+ P_{10}H_D+P_{11}H_I\;, \;\;\;\label{decomp}
\ee
with projectors $P_{ij}$ on the respective states (with $\sum_{ij=0}^1 P_{ij}=1$).
Each Hamiltonian $H_k$ ($k={\rm C},{\rm D}, {\rm M}, {\rm I})$ is written in terms of an energy matrix $F^{(k)}$ just as in Eq.~(\ref{hamc})
\be
F^{(C)}&=&\bordermatrix{\mbox{} & {\rm C}& {\rm D} & {\rm M} & {\rm I} \cr
                            {\rm C}   & 0  & \frac r3 & 0 & \frac r3 \cr
                            {\rm D}     &\frac r3 &  \frac23r  & \frac r3 &  \frac23r\cr
                            {\rm M} & 0  & \frac r3 & 0 & \frac r3 \cr
                            {\rm I} &\frac r3 &  \frac23r  & \frac r3 &  \frac23r\cr} 
                         =   \left(\begin{array}{cc}
                					E^{(C)}  & E^{(C)} \cr
                            E^{(C)} &  E^{(C)}\end{array}\right)
                            \label{ec} \;.\;\;\;\;
                            \ee
\begin{figure}[htbp] 
   \centering
   \includegraphics[width=3in]{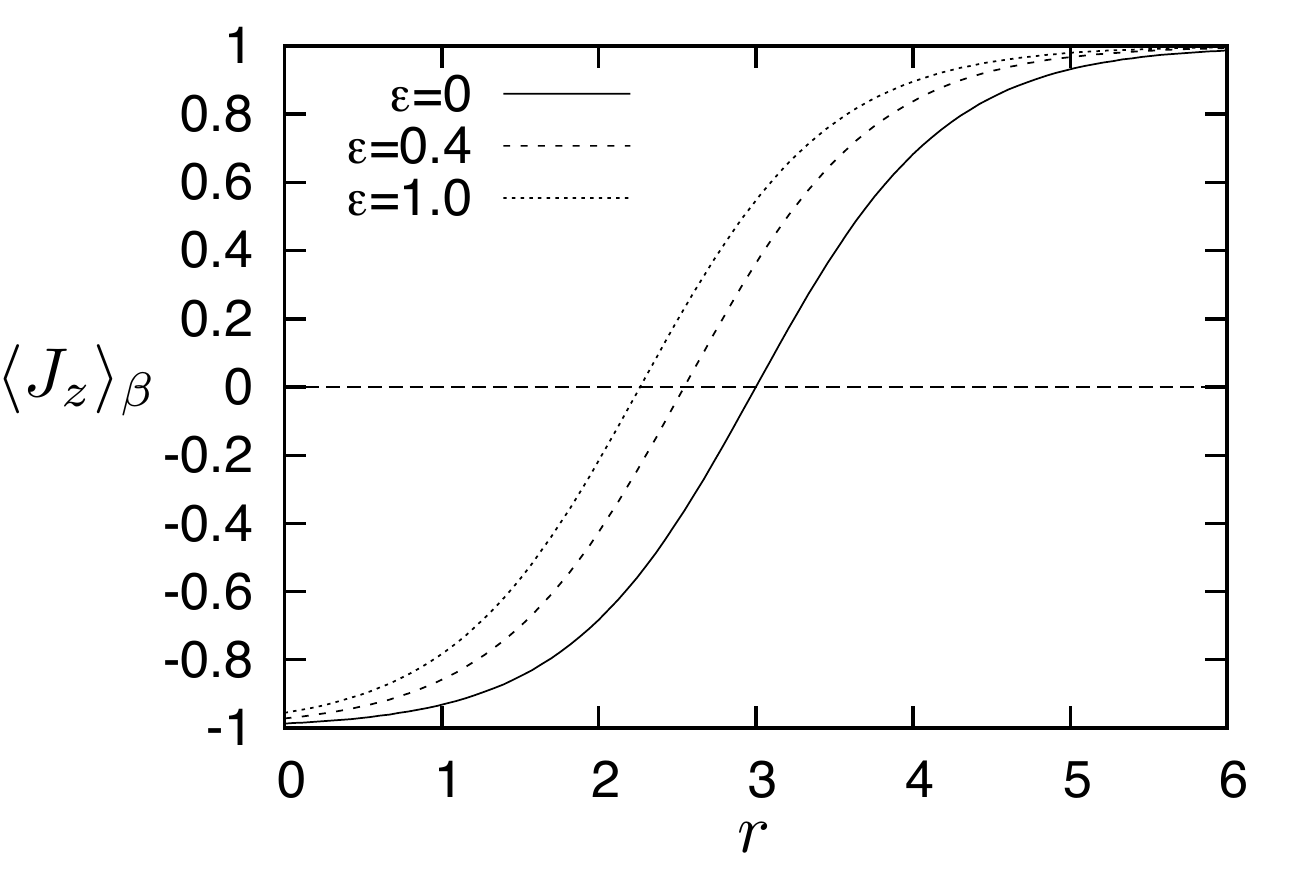} 
   \caption{Exact results for the order parameter $\la J_z\ra_\beta$ for the Public Goods game with punishment, as a function of synergy parameter $r$ for three different punishment fines $\epsilon$, at a constant temperature ($\beta=5$).}
   \label{fig:punish}
\end{figure}                            
Similarly,
\be                            
F^{(D)}&=&\frac r3-1+     \left(\begin{array}{cc}
                					E^{(C)}  & E^{(C)}+\frac\epsilon2 \cr
                            E^{(C)}+\frac\epsilon2 &  E^{(C)}+\epsilon\end{array}\right)                       \;, \label{ed} \\
F^{(M)}&=&                            \left(\begin{array}{cc}
                					E^{(C)}  & E^{(C)}+\frac\gamma2 \cr
                            E^{(C)}+\frac\gamma2 &  E^{(C)}+\gamma\end{array}\right) \;,       \label{em} \\      
 F^{(I)}&=&\frac r3-1+     \left(\begin{array}{cc}
                					E^{(C)}  & E^{(C)}+\frac{\gamma +\epsilon}2 \cr
                            E^{(C)}+\frac{\gamma+\epsilon}2&  E^{(C)}+\gamma+\epsilon\end{array}\right).  \;           \;\;      \;\;    \; \label{ei} \\    \nonumber               
 \ee
We can now calculate the partition function
\be
Z=\tr (e^{-\beta H})=Z_{\rm C}+Z_{\rm D}+Z_{\rm M}+Z_{\rm I}
\ee
on account of the decomposition (\ref{decomp}), where
\be
Z_{\rm C}=\tr(e^{-\beta H_C})=4(1+e^{-\beta\frac r3})^2\;,
\ee
or four times the contribution from each $E^{(C)}$. Similarly,
\be
Z_{\rm D}&=&e^{-\beta(\frac13 r-1)}(1+e^{-\beta\frac r3})^2(1+e^{-\beta\frac\epsilon2})^2 \;,\\
Z_{\rm M}&=&4(1+e^{-\beta(\frac r3+\frac\gamma2)})^2\;,\\
Z_{\rm I}&=&e^{-\beta(\frac13 r-1)}(1+e^{-\beta(\frac r3+\frac\gamma2})^2(1+e^{-\beta\frac\epsilon2})^2\;.\;\;\;\;\;\;\;
\ee
Finally, we obtain the order parameter that measures the degree of cooperation (the fraction of C and M players minus the fraction of D and I players), which turns into the surprisingly simple expression
\be
\la J_z\ra_\beta=\frac{1-\cosh^2(\beta\frac\epsilon4)e^{-\beta(\frac r3+\frac\epsilon2-1)}}{1+\cosh^2(\beta\frac\epsilon4)e^{-\beta(\frac r3+\frac\epsilon2-1)}} \label{solpunish}\;.
\ee
Note that the order parameter only depends on the effect of punishment $\epsilon$ but {\em not} the cost $\gamma$, and reduces to expression (\ref{order}) in the limit $\epsilon\to0$. 

\begin{figure}[htbp] 
   \centering
   \includegraphics[width=3in]{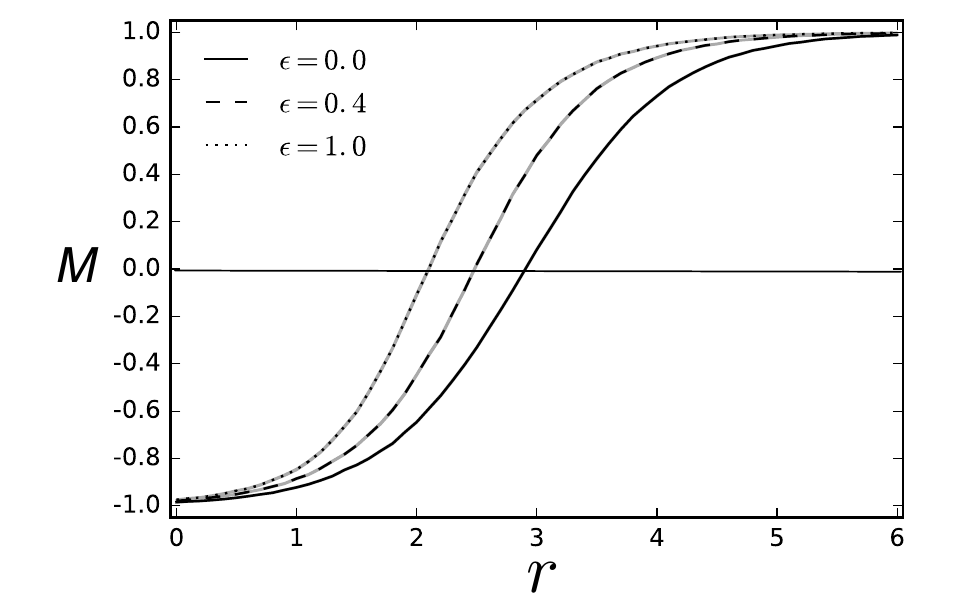} 
   \caption{Fraction of cooperators $M$ for a chain of length $2^{10}$ as a function of the synergy parameter $r$ for three different fines $\epsilon$ (at fixed $\beta=5$). Each data point (increments of $\delta r=0.1$) is the average over 100 replicates running the agent-based simulation with Glauber dynamics for $2\times 10^6$ updates.}
   \label{ABS_PGG_withPunish}
\end{figure}

To check the theory, we can extend the agent-based model described above by including the two new strategies I and M. As before, we use 1024 players in a population that is arranged linearly (see Methods), and games are played in groups of three. Again, when we evolve this population using strategy adoption, we see the dependence of the critical point on the synergy factor $r$ and the selection strength $\beta=1/T$. 
Since the game now includes two more strategies, we have to modify the function $M$ that describes the fraction of cooperators in the game  to contain all four strategies as the fraction of contributing (cooperating)  strategies:
\be
M=\frac{(P_{\rm C}+P_{\rm M})-(P_{\rm D}+P_{\rm I})}{P_{\rm C}+P_{\rm D}+P_{\rm M}+P_{\rm I}} \label{orderParameterPGGsimulationWithPunishment}
\ee
Evolving these populations using different fines $\epsilon$ and costs $\gamma$, we find that the critical point now only depends on $\epsilon$ (see Figure~\ref{ABS_PGG_withPunish}), and moves the critical point in such a manner that the punishment fine reduces the critical synergy for cooperation~\cite{HintzeAdami2015}.
It turns out that the closed-form solution Eq.~(\ref{solpunish}) reproduces the agent-based simulations shown in Fig.~\ref{ABS_PGG_withPunish} to a remarkable extent, confirming the unintuitive finding that the critical point only depends on the effect, but not on the cost, of punishment. The Hamiltonian model also clarifies that punishment indeed acts like a magnetic field that encourages alignment of spins, thus explaining why in agent-based simulations punishment induces hysteresis as a population is subjected to an adiabatically varying $r$~\cite{HintzeAdami2015}. Further work using the Hamiltonian model of cooperation with punishment may elucidate other aspects of the critical dynamics, in particular for games in higher dimensions, with more players per group, or even on irregular lattices.

\subsection{Discussion}
Evolutionary Game Theory is a mathematical framework that has been eminently successful at unraveling the numerous elements that impact decisions, and to work out the decision's consequences. While both mathematics and computational simulations have influenced this field (see for example the review~\cite{Adamietal2016a}, along with commentaries), the relationship between game theory and physics has been explored less. In real situations, decisions must be made under uncertainty; either due to unpredictable environments, or due to inherent noise. For evolutionary dynamics in particular, noise is unavoidable. After all, high reproductive potential does not guarantee survival, but only biases future outcomes. A standard result of population genetics for example predicts that a gene that confers a ten percent advantage in reproductive rate only has a twenty percent chance of being represented in future generations. The branch of science best equipped to tackle the impact of chance on dynamics is physics, with a well-developed corpus of results in statistical mechanics and thermodynamics. A growing literature has found success in mining these well-established methods, from harnessing the Fokker-Planck equation to describe the effect of chance due to drift in small populations~\cite{TraulsenHauert2009} to using tools from statistical mechanics to study the universality class of phase transitions in the spatial Prisoner's Dilemma~\cite{HauertSzabo2005}. Here, we tapped a different set of well-established tools from statistical physics, namely the thermodynamics of spin systems. The analogy between the critical dynamics of spin systems and game theory is not difficult to see. After all, the correspondence between Eigen and Schuster's model for the evolution of macromolecules~\cite{EigenSchuster79} and two-dimensional Ising models was pointed out over thirty years ago~\cite{Leuthaeusser1987} (see also section 11.4 in ~\cite{Adami1998})
but we have not, as yet, seen a concerted effort to marshal the considerable machinery developed to tackle low-dimensional condensed matter structures to aid in understanding evolutionary game theory. 

It may seem odd, at first sight, that a thermodynamic approach to game theory is possible at all, given that thermodynamics relies on the assumption that the system tends towards equilibrium, whereas in many game-theoretic situations (in particular, those that are of the Rock-Paper-Scissors type) the system appears to be maintained out of equilibrium. Fortunately, it is possible to show that even in systems out of equilibrium, detailed balance can be assured as long as microscopic reversibility is guaranteed~\cite{GrahamHaken1971,Risken1972}. While this result depends on the nature of the boundary condition (it holds under ``normal" boundary conditions, that is, boundary conditions in which the probability distribution vanishes at the boundary), there are strong reasons to believe that at least in the limit of large systems and low mutation rates, detailed balance can always be achieved for these games. Investigating this issue more deeply is left for forthcoming work.

The Hamiltonian approach we described here leads to important new insights about the dynamics of evolving populations at fixed strength of selection (and thus, to some extent, fixed temperature). First, we have shown that the standard statistical approach in which the energy of the entire ensemble is minimized, does not correspond to the evolutionary scenario, giving rise instead to a transition at $r=1$. That result would imply that a dilemma is absent, and indeed this is precisely what we would expect if groups of organisms, rather than individuals, are selected.
Second, the treatment of the Public Goods Game with punishment revealed that punishment plays the role that a local magnetic field plays when interacting with a system that can display spontaneous magnetization. An extensive literature in the area of spin-glasses of the Sherrington-Kirkpatrick type~\cite{SherringtonKirkpatrick1975} suggests that 
local magnetic fields can give rise to spontaneous symmetry breaking, and that their mean-field solutions are similar to those of local spins interacting with a global magnetics field. These insights immediately suggests to look for effects such as hysteresis (as seen, for example, in~\cite{HintzeAdami2015}), but also interactions between hysteresis and impurities for example. Indeed, it is not unreasonable to imagine that including a third player strategy such as ``abstaining"~\cite{Fowler2005,Brandtetal2006,Hauertetal2008} can be viewed as impurities that can dramatically alter the critical dynamics we observe, for example by ``pinning" the interfaces between domains\footnote{
Note that for this analogy to hold, the fraction of abstaining players (the density of impurities) must be kept constant. In general, these impurities can move through the crystal, but were kept fixed in numerical simulations of this effect~\cite{VainsteinArenzon2001}}. While the dynamics that includes abstaining may give rise to intransitive dominance dynamics~\cite{Hauertetal2002} (such as the Rock-Paper-
Scissors game), the arguments given above that out-of-equilibrium dynamics still gives rise to stationary equilibrium distributions as long as microscopic reversibility holds, suggest that such games can also be solved using Hamiltonian dynamics. 

We should caution, however, that extending the present results to games in higher dimensions will be difficult. For example, while the Ising model can be solved in two dimensions, there is no solution for the model in two dimensions with a magnetic field, as it is related to the three-dimensional model for which a closed-form solution does not exist.  Nevertheless, we expect that the tools developed here will be useful because if the analogy between evolutionary game dynamics and phase transitions in spin systems is established, other results from the rich literature of critical phenomena in spin systems may inform us about the dynamics of cooperation in groups.  In particular, an extension of the calculation shown here to two dimensions may produce an exact solution along the lines of Onsager's, which would allow us to move beyond pair-approximations for games on a 2D regular lattice. We hope that the simple results derived here (validated via computational simulation) can serve as a seed for the future development of this field. 

\section*{Methods}
The computational evolutionary model instantiates a population of 1,024 random agents in a circular configuration. At each update a single agent is randomly selected and its payoff computed by playing the  strategy against its left and right neighbors. At the same time, the payoff of a strategy to potentially replace the agent is computed. In case of a two player game (C and D) the only other alternative strategy is used, in the case of four players (C, D, M, and I) one alternative strategy is chosen at random. Instead of the evolutionary updating of the population described in~\cite{HintzeAdami2015,Adamietal2016a}, here the likelihood to replace the strategy of the selected agent with the alternative is given by Eq.~(\ref{strategyAdoptionProbability}). In each replicate run, we updated strategies 2 million times (roughly 2,000 updates per site), then calculated the order parameter. The code as well as the analysis scripts to create all figures can be found at: \textit{github reference will be provided upon acceptance of the manuscript}.

\section*{Acknowledgements} We thank Nathaniel Pasmanter for collaboration in the early stages of this work, as well as Claus Wilke for discussions. This work was supported in part by NSF's BEACON Center for the Study of Evolution in Action, under Contract No. DBI-0939454.  We wish to acknowledge the support of the Michigan State University High Performance Computing Center and the Institute for Cyber-Enabled Research.

\end{document}